\documentclass{IEEEtran}

\usepackage{amsmath}
\usepackage{amssymb}
\usepackage{cite}
\newcommand{\bs}{}

\newtheorem{theorem}{Theorem}
\newtheorem{remark}{Remark}
\newtheorem{asum}{Assumption}

\begin{document}

\title{  Total Energy Shaping of Underactuated Systems via Gravity Compensation at the Desired Equilibrium}

\author{ M. Reza J. Harandi and Mehrzad Namvar
%	\thanks{This work is based upon research funded by Iran National Science Foundation
%		(INSF) under project No.4031406.}
% \thanks{M.~Reza~J.~Harandi  is
% 	 with the Department of Electrical, Computer Engineering, Qom University of Technology, P.O. Box 3718146645, Qom, Iran (e-mail: jafariharandi@qut.ac.ir).}
% 	\thanks{ Mehrzad Namvar is with Department of Electrical Engineering, Sharif University of Technology, Tehran, Iran (email: namvar@sharif.ir)}
}
\maketitle

\begin{abstract}
This paper investigates total energy shaping of underactuated mechanical systems with gravity compensation evaluated at the desired equilibrium instead of throughout the online closed-loop operation. Within the interconnection and damping assignment passivity-based control (IDA-PBC) framework, sufficient conditions for local asymptotic stability are established in terms of integrability of a vector field and positive definiteness of an augmented potential energy. The approach is also combined with a reformulation of potential energy matching that replaces a nonhomogeneous partial differential equation with a homogeneous one and introduces an additional design term. Analytic constructions for the TORA system and a class of two-degree-of-freedom systems illustrate how the matching and integrability requirements can be addressed. The proposed control law avoids online evaluation of the configuration-dependent gravity term; offline controller synthesis and verification can still require knowledge of the gravity model.
\end{abstract}

\section{Introduction}\label{s1}
PD control with gravity compensation is a well-established approach for stabilizing fully actuated mechanical systems~\cite[ch. 3.1]{ortega2013passivity}. A simplified variant, referred to as PD with desired gravity compensation (DGC), compensates gravity only at the desired equilibrium point (EP) rather than along the entire trajectory~\cite{kelly1997pd}. Such PD designs do not directly address general underactuated mechanical systems (UMSs), for which IDA-PBC (interconnection and damping assignment passivity-based control) provides a total-energy-shaping framework~\cite{acosta2005interconnection}. In conventional IDA-PBC implementations, the input may depend on the configuration-dependent gravity term. This work studies an alternative implementation that uses only the gravity value at the desired EP online. The associated offline design conditions, however, can depend on the full potential energy.

PD with DGC can be interpreted as a potential energy shaping method in which the gravity term, given by the gradient of the potential function, is compensated only at the desired EP. Its stability relies on the boundedness of the gravity differential and requires a suitable condition on the controller gain; see~\cite{harandi2023practical,zavala2007natural} for details. In contrast, IDA-PBC provides a general framework for stabilizing UMSs via total energy shaping. It constructs the desired closed-loop dynamics in port-Hamiltonian form~\cite{yaghmaei2023contractive} and determines admissible energy functions by solving matching conditions in the form of partial differential equations (PDEs)~\cite{7393499}. Various works have addressed the solution or simplification of these PDEs for specific classes of systems~\cite{harandi2023reformulation,romero2018global,harandi2022solution,ortega2021pid,harandi2021matching,donaire}.

Integrating total energy shaping with gravity compensation applied only at the desired EP introduces additional structural conditions, most notably the integrability of a vector field depending on the system dynamics and the selected desired inertia matrix, together with a positive definiteness condition analogous to that in PD with DGC~\cite{kelly1997pd}. Unlike standard IDA-PBC, where the full gravity contribution is embedded in the shaped potential energy, the proposed approach compensates gravity only at the desired EP, resulting in a modified closed-loop structure.

In this paper, the IDA-PBC framework is modified accordingly, and sufficient conditions guaranteeing asymptotic stability of the desired EP are established. Moreover, the integrability requirement is effectively relaxed by combining the proposed approach with the method in~\cite{harandi2023reformulation}, which avoids solving the full potential energy PDE and introduces additional degrees of freedom for controller design. Conditions relevant to local stability and the construction of Lyapunov sublevel sets are discussed, and classes of systems admitting simplified controller synthesis are identified through analytic examples.

The main contributions are summarized as follows:
\begin{itemize}
	\item An IDA-PBC formulation for UMSs in which the gravity term is evaluated at the desired equilibrium in the online control law.
	\item Sufficient conditions for asymptotic stability based on vector-field integrability and a positive definiteness condition.
	\item Integration with the IDA-PBC reformulation in \cite{harandi2023reformulation}, avoiding the full potential energy PDE and introducing additional degrees of freedom that facilitate integrability.
	\item Analytic examples and remarks identifying cases in which the integrability conditions can be satisfied or reduced to ordinary differential equations.
\end{itemize}

The objective is to characterize conditions under which the desired-gravity-compensation philosophy can be combined with total energy shaping, rather than to claim universal superiority over conventional IDA-PBC. In particular, the reduced online gravity evaluation should be distinguished from the information needed during offline synthesis and analysis. The stability guarantees established below are local unless additional invariant-set conditions are verified.

	\section{Preliminaries}\label{sec:2}
For completeness, we briefly review IDA-PBC and PD with DGC; see the cited references for details.
	\subsection{IDA-PBC framework for UMSs}\label{sec:21}
Here, IDA-PBC for UMSs is briefly reviewed~\cite{harandi2021matching,acosta2005interconnection}. The equations governing the motion of a simple mechanical setup expressed through the port-Hamiltonian approach are given by
	\begin{equation}
	\label{1}
	\begin{bmatrix}
	\dot{q} \\ \dot{p}
	\end{bmatrix}
	=\begin{bmatrix}
	0_{n\times n} & I_n \\ -I_n & -R(q)
	\end{bmatrix}
	\begin{bmatrix}
	\nabla_q H \\ \nabla_p H
	\end{bmatrix}
	+\begin{bmatrix}
	0_{n\times m} \\ G
	\end{bmatrix}
	u
	\end{equation}
	in which $q,p\in \mathbb{R}^n$ represent the position and momentum, respectively, with $p=M\dot{q}$. $H(q,p)=V(q)+\frac{1}{2}p^TM^{-1}(q)p$ represents the total energy, consisting of potential and kinetic energy.  $M(q)$ is
	the positive definite inertia matrix, $R(q)=R^T(q)\succeq 0$ is the natural damping matrix, and the full-column-rank input matrix is denoted by $G(q)\in\mathbb{R}^{n\times m}$, where $m<n$. Considering the following control law
			\begin{align}
	&u=(G^TG)^{-1}G^T\big(\nabla_q H-M_dM^{-1}\nabla_q H_d+J_2\nabla_p H_d\big)\nonumber\\&-K_vG^T\nabla_p H_d,
	\label{4}
	\end{align}
where the following matching equations determine $M_d$ and $V_d$
		\begin{align}
	&G^\bot \{\nabla_q  \big(p^TM^{-1}(q)p\big) - M_dM^{-1}\nabla_q \big(p^TM_d^{-1}(q)p\big)\nonumber\\   & +2J_2M_d^{-1}p \} =0_{(n-m)\times 1},   \label{5} \\&
	G^\bot \{\nabla_q V-M_dM^{-1}\nabla_q V_d\} =0_{(n-m)\times 1}, \label{6}
	\end{align}
	where $G^\bot(q)\in \mathbb{R}^{(n-m)\times n}$ is a full-rank left annihilator of $G(q)$, i.e.,
$	G^\bot(q)G(q)=0,$
	and $\mathrm{rank}(G^\bot)=n-m$.
Accordingly, the closed-loop equations are formulated as
		\begin{equation}
	\label{2}
	\begin{bmatrix}
	\dot{q} \\ \dot{p}
	\end{bmatrix}
	=
	\begin{bmatrix}
	0_{n\times n} & M^{-1}M_d \\ -M_dM^{-1} & J_2-R_2
	\end{bmatrix}
	\begin{bmatrix}
	\nabla_q H_d \\ \nabla_p H_d
	\end{bmatrix}.
	\end{equation}
Here, $J_2(q,p)\in\mathbb{R}^{n\times n}$ is skew-symmetric,
$R_2=RM^{-1}M_d+GK_vG^T$, and $K_v=K_v^T\succ0$ is the damping gain. Since $R_2$ need not be symmetric, write $D_2=(R_2+R_2^T)/2$ and assume $D_2\succeq0$ in the stability arguments; see also~\cite{franco2026passivity}.
	The desired equilibrium $q^*$ must satisfy $G^\bot\nabla_q V(q^*)=0$. In the conventional IDA-PBC design, $V_d$ is selected to have an isolated minimum at $q^*$.
Under the stated dissipation and potential conditions, the Lyapunov derivative satisfies
		\begin{align}
		\dot{H}_d=-(\nabla_p H_d)^TD_2\nabla_p H_d\leq0. \label{vdot}
		\end{align}
		Local asymptotic stability follows if $D_2\succ0$, or under an appropriate local zero-state detectability condition for the dissipation output $D_2^{1/2}\nabla_pH_d$.
	Note that as customary in the IDA-PBC literature, we assume that the matching equations (\ref{5}) and (\ref{6}) admit a solution.
	\subsection{Review of PD with DGC for fully actuated systems}\label{sec:22}
	In this case, the term $\nabla_{q}V$ in the PD controller with gravity compensation is replaced by $\nabla_{q}V|_{q=q^*}$, where $\nabla_q V$ denotes the gravity term~\cite{kelly1997pd}. Accordingly, the control law of PD with desired gravity compensation becomes
$u=\nabla_{q}V|_{q=q^*}-K_p(q-q^*)-K_d\dot{q},$
in which $\nabla_{q}V|_{q=q^*}$ is gravity term at desired EP, and $K_p$ and $K_d$ are positive definite gains.
The desired equilibrium point is asymptotically stable if
 	\begin{align}\label{gg}
\lambda_{min}\{K_p\}>\kappa_g,
	\end{align}
in which $\kappa_g$ is a constant satisfying
	\begin{align}\label{ggg}
		\|\nabla_{q}V(q_1)-\nabla_{q}V(q_2)\|\leq \kappa_g\|q_1-q_2\|,
			\end{align}
for every $q_1$ and $q_2$ in the workspace of the robot~\cite{kelly1997pd,harandi2023practical}.

Motivated by this observation, the next section extends the DGC idea to UMSs within an energy shaping framework.

	\section{IDA-PBC with Gravity Compensation at the Desired Equilibrium Point}\label{sec:3}
Designing IDA-PBC with gravity compensation applied only at the desired equilibrium point $q^*$ requires distinguishing between actuated and unactuated CaVs. Accordingly, the following assumption is imposed.
\begin{asum}\label{as1}\normalfont
The input matrix is a constant coordinate-selection matrix, of the form $G=P[I_m,0_{m\times (n-m)}]^T$ in which $P\in\mathbb{R}^{n\times n}$ represents a permutation matrix.$\hfill\square$
\end{asum}
This assumption is satisfied by numerous benchmark mechanical models with joint torques as inputs; it excludes general configuration-dependent input mappings (see, e.g.,~\cite{harandi2023stabilization,acosta2005interconnection}).
Under \textit{Assumption~\ref{as1}}, the actuated ($q_a$) and unactuated ($q_u$) CaVs are defined as
 \begin{align*}
[q_a , q_u]^T
:=P^Tq,\qquad q_a\in\mathbb{R}^{m},q_u\in\mathbb{R}^{n-m}.
\end{align*}
The corresponding control law is given by
\begin{align}
&u=u_{IDA}-\nabla_{ q_a}V+\left.\big(\nabla_{ q_a}V\big)\right|_{ q= q^*}=( G^T G)^{-1} G^T\nonumber\\&\big(\nabla_{{q}} K+\nabla_{{q}} V|_{q=q^*} -{M_d}{M}^{-1}\nabla_{{q}} H_d+{J_2}{M_d}^{-1} p\big)\nonumber\\&-{K_v} G^T\nabla_{{p}} H_d,\label{u}
\end{align}
where $u_{IDA}$ denotes the IDA-PBC control law in (\ref{4}) and $K$ is the kinetic energy. Note that in (\ref{u}), unlike (\ref{4}), the gravity term is compensated only at the desired EP.
Under (\ref{5}) and (\ref{6}), the closed-loop dynamics become
     \begin{align}
\begin{bmatrix}
\dot{\bs q} \\ \dot{\bs p}
\end{bmatrix}
&=\begin{bmatrix}
0_{n\times n} & \bs M^{-1}\bs{M_d} \\ -\bs{M_d}\bs M^{-1} & J_2-R_2
\end{bmatrix}
\begin{bmatrix}
\nabla_{\bs{q}} H_d \\ \nabla_{\bs{p}} H_d
\end{bmatrix}\nonumber\\&+\begin{bmatrix}
0_{n\times 1} \\ G(G^TG)^{-1}G^T(\nabla_{{q}} V|_{q=q^*}-\nabla_q V)
\end{bmatrix},\label{cl1}
\end{align}
where the last term indicates that gravity remains in the actuated coordinates.
Following \cite{harandi2023reformulation}, suppose that (\ref{6}) is replaced by the PDE
\begin{align}\label{pd}
G^\bot M_dM^{-1}\nabla_q V_d =0_{(n-m)\times 1}.
\end{align}
The closed-loop system's dynamic behavior is  described by
    \begin{align}
\begin{bmatrix}
\dot{\bs q} \\ \dot{\bs p}
\end{bmatrix}
&=\begin{bmatrix}
0_{n\times n} & \bs M^{-1}\bs{M_d} \\ -\bs{M_d}\bs M^{-1} & J_2-R_2
\end{bmatrix}
\begin{bmatrix}
\nabla_{\bs{q}} H_d \\ \nabla_{\bs{p}} H_d
\end{bmatrix}\nonumber\\&+\begin{bmatrix}
0_{n\times 1} \\ \nabla_{\bs q} V|_{\bs q=\bs q^*}-\nabla_{\bs q} V
\end{bmatrix},\label{cl2}
\end{align}
where
$\nabla_{\bs q_u} V|_{\bs q=\bs q^*}=(G^\bot \nabla_{{q}}V)|_{\bs q=\bs q^*}=0$ has been used~\cite{harandi2023reformulation}. Note that, similar to IDA-PBC, deriving the closed-loop system (\ref{cl1}) and (\ref{cl2}) is straightforward by multiplying the dynamics of $p$ in (\ref{1}) with $[G,G^{\bot^T}]^T$ while taking into account the control law and the PDEs.
In general, neither $H_d$ nor the Lyapunov function used in~\cite{kelly1997pd} is sufficient to establish stability of (\ref{cl1}) or (\ref{cl2}). This difficulty arises because the gravity term is not fully shaped into the closed-loop energy. Consequently, additional structural conditions are required, as formalized in the following theorem.

\begin{theorem}\normalfont\label{t}
Consider IDA-PBC with gravity compensation at $q^*$ given by (\ref{u}), under (\ref{5}) and \textit{Assumption~\ref{as1}}. Assume $D_2\succeq0$ and either $D_2\succ0$ or local zero-state detectability of $D_2^{1/2}\nabla_pH_d$. Suppose further that $\nabla_qV_d(q^*)=0$. Then $(q^*,0)$ is locally asymptotically stable if one of the following cases holds:

		\textbf{a})
		  The potential function is separable, i.e., $V(\bs q)=V_a(\bs q_a)+V_u(\bs q_u)$, and (\ref{6}) is satisfied.
		The vector
		\begin{align}
	\bs\eta_1(\bs q):=\bs M(\bs q) \bs{M_d^{-1}}(\bs q)\Big( \nabla_{\bs q}V_a(\bs q_a)-\nabla_{\bs q} V_a(\bs q_a)|_{q=q^*}\Big) ,
		\label{p13}
		\end{align}
	is integrable, i.e., there exists a function $\phi_1(q)$ such that $\nabla_{q}\phi_1=\eta_1$, and
		\begin{align}
		\left.\Big(	\frac{\partial^2V_d}{\partial \bs q^2}+\bs{MM_d^{-1}}\frac{\partial^2V_a}{\partial \bs q^2}\Big)\right|_{q=q^*}> 0.\label{p14}
		\end{align}

\textbf{b}) The PDE (\ref{pd}) is satisfied, and the vector
\begin{align}
\bs\eta_2(\bs q):=\bs M(\bs q) \bs{M_d^{-1}}(\bs q)\Big( \nabla_{\bs q}V(\bs q)-\nabla_{\bs q} V(\bs q)|_{q=q^*}\Big) ,
\label{p8}
\end{align}
is integrable, i.e., there exists a function
$\phi_2(q)$ such that
$\nabla_q\phi_2=\eta_2,$ and
\begin{align}
\left.\Big(	\frac{\partial^2V_d}{\partial \bs q^2}+\bs{MM_d^{-1}}\frac{\partial^2V}{\partial \bs q^2}\Big)\right|_{q=q^*}> 0.\label{p12}\quad \square
\end{align}

\end{theorem}
 The integrability conditions can be restrictive. The cases in Section~\ref{s4} and Remark~\ref{r1} illustrate circumstances in which they simplify. The theorem is local; no equality of attraction regions with standard IDA-PBC is implied.

\begin{proof}
	\textbf{a})The closed-loop dynamics are given by (\ref{cl1}). Since $G(G^TG)^{-1}G^T\nabla_q V_a=\nabla_q V_a$ and $G(G^TG)^{-1}G^T\nabla_q V_u=0$, the last term in (\ref{cl1}) reduces to $[0_{n\times 1} , \nabla_{{q}} V_a|_{q=q^*}-\nabla_q V_a]^T$.
Consider the Lyapunov candidate
$	\mathcal{V}:=H_d+\phi_1.$
	Let $v=\nabla_pH_d$ and
$\Delta_a(q)=\nabla_qV_a(q_a)-\nabla_qV_a(q_a^*)$.
Since $\dot q=M^{-1}M_dv$ and
$\nabla_q\phi_1=MM_d^{-1}\Delta_a$, differentiation along
(\ref{cl1}) gives
\begin{align*}
\dot{\mathcal V}
&=-v^TR_2v-v^T\Delta_a
  +(M^{-1}M_dv)^TMM_d^{-1}\Delta_a\\
&=-v^TD_2v\leq0.
\end{align*}
It remains to verify that $\mathcal V$ is a valid Lyapunov function. Since $\eta_1$ is integrable, $\mathcal V$ is well defined and positive definite with respect to $p$. Therefore, it suffices to examine the $q$-dependent terms of $\mathcal V$, denoted by $\mathcal V'$.
Accordingly, it suffices to verify
	\begin{align}
	\left.({\partial \mathcal{V}'}/{\partial q})\right|_{q=q^*}=0,\qquad \left.({\partial^2 \mathcal{V}'}/{\partial q^2})\right|_{q=q^*}>0.\label{con}
	\end{align}
The first condition in (\ref{con}) holds since $\mathcal{V}'=V_d+\phi_1$ and
	\begin{align*}
\nabla_{{q}}\mathcal{V}'|_{q=q^*}=	\left.\nabla_q V_d\right|_{q=q^*}+\bs M(\bs q^*) \bs{M_d^{-1}}(\bs q^*)&\\\Big( \nabla_{\bs q}V_a(\bs q_a)|_{q=q^*}-\nabla_{\bs q} V_a(\bs q_a)|_{q=q^*}\Big)=0,&
	\end{align*}
	since $\nabla_qV_d(q^*)=0$.
The Hessian in the second condition of (\ref{con}) becomes
		\begin{align*}
	&\nabla_{{q}}^2V_d|_{q=q^*}+\bs M(\bs q^*) \bs{M_d^{-1}}(\bs q^*) \nabla_{{q}}^2V_a|_{q=q^*}\\&+\nabla_q(MM_d^{-1})|_{q=q^*}\Big( \nabla_{\bs q}V_a(\bs q_a)|_{q=q^*}-\nabla_{\bs q} V_a(\bs q_a)|_{q=q^*}\Big),
	\end{align*}
	which is positive definite by (\ref{p14}).Hence, $\mathcal V$ is a valid Lyapunov function and stability of $q^*$ follows. Local asymptotic stability follows from the specified strict dissipation or detectability condition and the local invariance principle.
	 Note that in this case, (\ref{cl1}) can be rewritten as
	  \begin{align}
		\begin{bmatrix}
			\dot{\bs q} \\ \dot{\bs p}
		\end{bmatrix}
		&=\begin{bmatrix}
			0_{n\times n} & \bs M^{-1}\bs{M_d} \\ -\bs{M_d}\bs M^{-1} & J_2-R_2
		\end{bmatrix}
		\begin{bmatrix}
			\nabla_{\bs{q}} \tilde{H}_d \\ \nabla_{\bs{p}} \tilde{H}_d
		\end{bmatrix},\label{cl}
	\end{align}
	in which $\tilde{H}_d=H_d+\phi_1$. Examples of systems with separable potential functions are reported in~\cite{romero2018global}.

	\textbf{b}) The proof follows the same steps as in part \textbf{a)}. By selecting $\mathcal{V}=H_d+\phi_2$, one obtains $\dot{\mathcal{V}}$ as in (\ref{vdot}), while (\ref{p12}) guarantees positive definiteness around $q^*$. Similarly, the closed-loop dynamics may be represented as (\ref{cl}) with $\tilde{H}_d=H_d+\phi_2$. $\hfill\blacksquare$

\end{proof}

In part \textbf{a)} of \textit{Theorem~\ref{t}}, a separable potential function was assumed. Nevertheless, the proposed controller can be applied to nonseparable potentials. In this case, $\eta_1$ is redefined as
		\begin{align*}
\eta_1(q)=\bs M \bs{M_d^{-1}}G(G^TG)^{-1}G^T\Big( \nabla_{\bs q}V-\nabla_{\bs q} V|_{q=q^*}\Big),
\end{align*}
and must be integrable. With $\nabla_q V_d(q^*)=0$, the condition
$	\left.({\partial^2 V_d}/{\partial \bs q^2}+{\partial\eta_1}/{\partial q})\right|_{q=q^*}> 0,$
should be satisfied. Since the proof follows the same steps as \textit{Theorem~\ref{t}}, it is omitted.
\begin{remark}\normalfont\label{r1}
The main limitation of the proposed approach is the integrability of the $\eta_i$ terms, which may restrict applicability of \textit{Theorem~\ref{t}}. However, in some cases, these conditions are automatically satisfied, e.g., potential energy shaping in which $M_d=M$ and
	$$\phi_1=V_a-q^T\nabla_{\bs q} V_a|_{q=q^*},\qquad \phi_2=V-q^T\nabla_{\bs q} V|_{q=q^*}.$$
	 UMSs characterized by $M=M(q_a)$ and $V=V_a(q_a)+V_u(q_u)$ provide another case where conditions (\ref{p13}) and (\ref{p8}) may be fulfilled using a diagonal desired inertia matrix. All diagonal elements are constant except those associated with the actuated coordinates, which depend on $q_a$ and can be designed directly from (\ref{5}) due to the vanishing forcing term; see, Section~\ref{tora} for an example.
Furthermore, when $M$ and $M_d$ are constant, a simpler condition may suffice. In this case, all the conditions of part \textbf{a)} and \textbf{b)} of \textit{Theorem~\ref{t}} remain unchanged, except that the functions $\eta_1$ and $\eta_2$ are replaced by
	$\bs\eta_3(\bs q):=\bs M \bs{M_d^{-1}}\nabla_{\bs q}V_a(\bs q_a),$
	and
	$\bs\eta_4(\bs q):=\bs M \bs{M_d^{-1}} \nabla_{\bs q}V(\bs q),$
	respectively. The proof follows directly from \textit{Theorem~\ref{t}}, and it is sufficient to consider
$\mathcal{V}=H_d+\phi_3-\bs q^T\bs M\bs M_d^{-1}\nabla_{\bs q}V_a(\bs q_a)|_{\bs q=\bs q^*},$
 and
  $\mathcal{V}=H_d+\phi_4-\bs q^T\bs M\bs M_d^{-1}\nabla_{\bs q}V(\bs q)|_{\bs q=\bs q^*},$
  as a Lyapunov functions for part \textbf{a)} and \textbf{b)} with $\phi_3$ and $\phi_4$ being the integral of $\eta_3$ and $\eta_4$, respectively. \hfill$\square$
\end{remark}
\begin{remark}\normalfont\label{r2}
	In \cite{harandi2023reformulation}, it was shown that under the integrability of the term
$$	MM_d^{-1}(G\xi+G^{\bot^T}G^\bot\nabla_q V),$$
	in which $-\xi(q)\in\mathbb{R}^m$ with $\xi(q^*)=0$ is a free term added to the control law,
	 it suffices to solve (\ref{pd}) instead of (\ref{6}). Combining the approach in~\cite{harandi2023reformulation} with part \textbf{b)} of \textit{Theorem~\ref{t}} further simplifies both controller design and implementation. The following theorem formalizes this result. \hfill$\square$
	\end{remark}
	\begin{theorem}\normalfont\label{t2}
Consider the control law
\begin{align}
&u=( G^T G)^{-1} G^T\big(\nabla_{{q}} K+\nabla_{{q}} V|_{q=q^*} -{M_d}{M}^{-1}\nabla_{{q}} H_d\nonumber\\&+{J_2}{M_d}^{-1} p\big)-{K_v} G^T\nabla_{{p}} H_d-\xi,\label{uu}
\end{align}
where $\xi(q^*)=0$. Assume $D_2\succeq0$ and either $D_2\succ0$ or local zero-state detectability of $D_2^{1/2}\nabla_pH_d$. Suppose that
	  \begin{align}\label{et}
	\eta_5(q):=\bs M \bs{M_d^{-1}}\left( \nabla_{\bs q}V-\nabla_{\bs q} V|_{q=q^*}+G\xi\right) ,
	  \end{align}
	 is integrable, i.e., there exists a function $\phi_5(q)$ such that $\nabla_{q}\phi_5=\eta_5$, and (\ref{5}) and (\ref{pd}) are satisfied. If $\nabla_qV_d(q^*)=0$, then $(q^*,0)$ is locally asymptotically stable provided that
	  \begin{align}\label{cn}
	  \left.\big(\frac{\partial^2 V_d}{\partial \bs q^2}+\frac{\partial\eta_5}{\partial q}\big)\right|_{q=q^*}> 0.\quad\square
	  \end{align}
	\end{theorem}

\begin{proof}
The closed-loop dynamics are identical to (\ref{cl2}) except for the last term, which is
	 $	 	[0_{n\times 1} , \nabla_{\bs q} V|_{\bs q=\bs q^*}-\nabla_{\bs q} V(\bs q)-G\xi]^T.$
It is straightforward to verify that
	 $	\mathcal{V}=H_d+\phi_5,$
	 is well defined and satisfies the local Lyapunov conditions in (\ref{con}) because $\eta_5(q^*)=0$ and (\ref{cn}) is positive definite. Moreover, $\dot{\mathcal{V}}$ is given by (\ref{vdot}), which completes the proof. $\hfill\blacksquare$
\end{proof}
\begin{remark}\normalfont\label{rr}
The Hessian conditions in Theorems~\ref{t} and~\ref{t2} certify a strict local minimum of the augmented potential, but do not by themselves establish a region of attraction. A larger invariant neighborhood may be certified by a compact sublevel set of $\widetilde H_d=H_d+\phi_i$ contained in the domain where $M_d\succ0$, $D_2\succeq0$, and the relevant detectability conditions hold, provided this sublevel set contains no other invariant zero-dissipation motions. Global claims additionally require properness of $\widetilde H_d$ and suitable global invariance conditions; properness of $H_d$ alone is insufficient. No general equality with the attraction region of conventional IDA-PBC is claimed. $\hfill\square$
\end{remark}

\textbf{Discussion:} The main challenge in applying the proposed approach lies in the integrability of the $\eta_i$ terms. However, as noted in \textit{Remark~\ref{r1}}, these conditions may be either simplified or entirely avoided in certain cases. In general, the $\eta_i$ terms depend on the system parameters $M$ and $V$, as well as on the design parameter $M_d$ derived from (\ref{5}).
Thus, the implementation of the proposed method is facilitated by a suitable choice of $M_d$, and this process is considerably simplified when the first term in (\ref{5}), commonly known as the forcing term, vanishes. Note that for some UMSs, the forcing term can be eliminated through coordinate transformations~\cite{viola2007total}.
 In particular, condition (\ref{et}) is generally easier to satisfy than the others because it contains the free term $\xi$. For some classes of UMSs, this flexibility allows the integrability condition to be reformulated as a set of ODEs, thereby simplifying controller synthesis despite the absence of a general closed-form solution; see Section~\ref{s42}.
\noindent\textbf{Relation to conventional IDA-PBC.} Whenever an integrable correction $\phi_i$ exists, the augmented energy $\widetilde H_d=H_d+\phi_i$ yields a port-Hamiltonian representation of the same closed-loop system. Thus the present construction is an alternative controller-synthesis and online-evaluation strategy within the IDA-PBC framework, not a claim of a distinct class of closed-loop dynamics. The integrability tests and the choice of $\xi$ may still use the full gravity model offline.

\section{Analytical Examples}\label{s4}
\subsection{TORA system}\label{tora}
The dynamic parameters of the Translational Oscillator with a Rotational Actuator (TORA) system are given in~\cite{olfati2001nonlinear} as
\begin{align*}
M=\begin{bmatrix}
a_1 & a_2\cos(q_2) \\ a_2\cos(q_2) & a_3
\end{bmatrix}, \hspace{.2cm} V=\frac{1}{2}k_1q_1^2-k_2\cos(q_2),
\end{align*}
and $G=[0,1]^T$ in which $q_1$ and $q_2$ denote the translational and rotational motions, respectively.
In order to stabilize $q^*=[0,q_{2d}]^T$, where $q_{2d}\in\mathbb{R}$, using IDA-PBC with gravity compensation at $q^*$, part \textbf{a)} of \textit{Theorem~\ref{t}} is employed. In this case, $V_a=-k_2\cos(q_2)$ and we choose $M_d=M$ which directly satisfies the integrability condition (\ref{p13}). The solution of (\ref{6}) is selected as
$V_d={1}/{2}\kappa(q_2-q_{2d})^2+{1}/{2}k_1q_1^2.$
Condition (\ref{p14}) becomes
\begin{align*}
	\begin{bmatrix}
		k_1 & 0 \\ 0 & \kappa+k_2\cos(q_{2d})
	\end{bmatrix}>0.
\end{align*}
Hence, $\kappa$ must satisfy
\begin{align}\label{ka}
	\kappa>-k_2\cos(q_{2d}).
\end{align}

To stabilize $q^*=[0,0]^T$, part \textbf{b)} of \textit{Theorem~\ref{t}} is employed. Following Remark~\ref{r1}, choose
$$M_d^{-1}=\begin{bmatrix}
\alpha & 0 \\ 0 & \beta(q_2)
\end{bmatrix},$$
then
$\eta_2^T=[\eta_{2_1},\eta_{2_2}]$ is integrable if \begin{align}\label{e}
\partial\eta_{2_1}/\partial q_2=\partial\eta_{2_2}/\partial q_1.
\end{align}
 This yields the following ODE
\begin{align*}
0.5k_2\sin(2q_2)[{d\beta}/{dq_2}]+k_2\cos(2q_2)\beta=k_1\alpha\cos(q_2),
\end{align*}
whose solution is $	\beta=(k_1 \alpha/k_2)\sec(q_2)$.
Moreover, (\ref{5}) is satisfied by selecting a skew-symmetric matrix $J_2$ with upper off-diagonal entry ${a_2\sin(q_2)p_2}/{(\alpha\det \bs M)}$.
The solution of the potential energy PDE (\ref{pd}) is selected as
 $$	V_d={1}/{2}\kappa\big(({a_2}/{a_3})q_1+\ln \left|{(1+\sin(q_2))}/{\cos(q_2)} \right|\big)^2.$$
Condition (\ref{p12}) becomes
 \begin{align*}
	\kappa	\begin{bmatrix}
		{a_2^2}/{a_3^2} & 	{a_2}/{a_3} \\ {a_2}/{a_3} & 1
	\end{bmatrix}
	+k_1\alpha\begin{bmatrix}
		a_1 & a_2 \\ a_2 & a_3
	\end{bmatrix},
\end{align*}
which is positive definite for $\kappa>0$.
This example illustrates how the reformulation approach
significantly facilitates the implementation of the proposed controller,
since the potential energy PDE (\ref{6}) is replaced by (\ref{pd}),
and the additional flexibility introduced through the selection of $M_d$
helps satisfy the integrability condition. It is worth noting that, for the selected nontrivial $M_d$, the standard potential energy PDE (\ref{6}) takes the form
\begin{align*}
	a_3\frac{\partial V_d}{\partial q_1}
	-
	a_2\cos(q_2)\frac{\partial V_d}{\partial q_2}
	=
	\alpha\big(a_1a_3-a_2^2\cos^2(q_2)\big)k_1q_1,
\end{align*}
whose nonhomogeneous term significantly complicates the direct solution of the PDE. In contrast, the proposed approach replaces (\ref{6}) with the homogeneous PDE (\ref{pd}), thereby simplifying the energy shaping procedure while still requiring gravity compensation only at $q^*$.

\subsection{ A Class of 2-DOF UMSs }\label{s42}

Consider a class of 2-DOF UMSs characterized by  $M=M(q_2), M_d=M_d(q_2)$ and $V=-k_1\cos(q_1)-k_2\cos(q_1+q_2)$ with $k_1,k_2>0$.  The goal is to illustrate the integrability conditions of \textit{Theorem~\ref{t2}} near a desired equilibrium $q^*$ satisfying $\nabla_qV(q^*)=0$ and $\nabla_qV_d(q^*)=0$. Related potential structures occur in benchmark systems such as the Acrobot and Pendubot~\cite{d2006further,harandi2021matching}; applicability to a particular model requires verifying its inertia and matching conditions.
Define
\begin{align*}
	MM_d^{-1}=\begin{bmatrix}
		a_1(q_2) & a_2(q_2) \\ a_3(q_2) & a_4(q_2)
	\end{bmatrix}.
\end{align*}
The vector $\eta_5^T$ becomes
\begin{align*}
	&[k_1a_1\sin(q_1)+k_2a_1\sin(q_1+q_2)+k_2a_2\sin(q_1+q_2)+a_i\xi,\\&k_1a_3\sin(q_1)+k_2a_3\sin(q_1+q_2)+k_2a_4\sin(q_1+q_2)+a_j\xi],
\end{align*}
where $(i,j)=(1,3)$ for $G=[1,0]^T$ and
$(i,j)=(2,4)$ for $G=[0,1]^T$.
Motivated by the structure of $\eta_5$, choose
\begin{align*}
	\xi=\gamma_1(q_2)\sin(q_1+q_2)+\gamma_2(q_2)\cos(q_1+q_2)&\\+\gamma_3(q_2)\cos(q_1)+\gamma_4(q_2)\sin(q_1)&,
\end{align*}
where the functions $\gamma_i(q_2)$ are to be designed. Applying the integrability condition of $\eta_5$, similarly to (\ref{e}), yields
\begin{align*}
	&k_2[{d(a_1+a_2)}/{dq_2}]+[{d(a_i\gamma_1)}/{dq_2}]+\gamma_2(a_j-a_i)=0,\\&
	[{d(a_i\gamma_2)}/{dq_2}]+k_2(a_1+a_2-a_3-a_4)+\gamma_1(a_i-a_j)=0,\\&
	[{d(a_i\gamma_3)}/{dq_2}]-k_1a_3-a_j\gamma_4=0,\\&
	k_1[{da_1}/{dq_2}]+[{d(a_i\gamma_4)}/{dq_2}]+a_j\gamma_3=0.
\end{align*}
For a specified admissible $M_d$, these equations provide an ODE characterization of the integrability requirement. Their local solvability, the condition $\xi(q^*)=0$, kinetic matching, and positivity of the augmented Hessian must still be checked; a closed-form solution is not guaranteed.

\section{Conclusion}
This paper formulated desired-equilibrium gravity compensation for underactuated mechanical systems within total energy shaping. Under matching, integrability, and augmented-energy positivity conditions, local asymptotic stability follows. The reformulation using a homogeneous potential matching equation and a free design term offers an alternative way to organize controller synthesis. The analytical TORA and two-degree-of-freedom constructions illustrate both the possibilities and the restrictions of the approach. Online evaluation of the configuration-dependent gravity term is avoided, whereas offline design and verification may still require the full model. Stronger attraction-region claims, systematic solvability results, and quantitative comparisons are left for future investigation.

\bibliographystyle{ieeetr}
	\bibliography{ref}

\end{document}